\documentclass[journal=nalefd,manuscript=letter]{achemso}
\usepackage{chemformula} 
\usepackage[T1]{fontenc} 
\usepackage{bm}
\usepackage{ragged2e}

\usepackage{chngcntr}

\usepackage{caption}
\author{Mingyue Wang}
\affiliation{Department of Physics, Xiamen University, Xiamen 361005, China}

\author{Jiayuan Wang}
\email{wangjiayuan@xmu.edu.cn}
\affiliation{Department of Physics, Xiamen University, Xiamen 361005, China}
\alsoaffiliation{Shenzhen Research Institute of Xiamen University, Shenzhen 518057, China} %
\title{Virtual Polarization Elements for Spatially Varying Jones Matrix Transformations on a Free-Space Plane}

\abbreviations{IR,NMR,UV} 

\begin{document}

\begin{abstract}
Precise control over the spatial and polarization properties of light is foundational for advanced photonic systems, yet most conventional approaches are constrained to local, contact-based manipulation at physical  interfaces. To overcome these constraints, here we introduce a fundamentally new framework for action-at-a-distance polarization control using virtual polarization elements (VPEs). VPEs apply prescribed local Jones matrix transformations between an input field at the modulation plane and an output field at a remote, contactless free-space plane, enabling polarization transformations without physical interaction at the target. We demonstrate VPEs, in metasurface platform, realizing diverse polarization functionalities, including single-function VPEs for circular polarizer, linear polarizer, half-wave plate, and quarter-wave plate operations; a multifunction VPE simultaneously implementing distinct polarization functions with arbitrary phase difference across spatial regions; and vortex waveplate configurations generating structured vector vortex beams. By decoupling the modulation and target planes, VPEs open new opportunities for remote polarization shaping, non-invasive beam engineering, and contactless polarization manipulation in challenging optical environments.
\end{abstract}

\section{Introduction}
The ability to precisely control the polarization state of light is foundational to modern optics and photonics, enabling applications in communications,\cite{martinelli2006polarization,zhao2009circle} sensing,\cite{yan2020general,tyo2006review} imaging,\cite{chenault2000polarization,demos1997optical} biomedical operation,\cite{novikova2012polarimetric,pierangelo2013polarimetric} quantum
optics,\cite{nape2022revealing,soderholm2012quantum} laser processing,\cite{porfirev2023laser,meier2007material} and all-optical computing.\cite{zhu2021topological,luo2022metasurface,zhu2017plasmonic,kwon2018nonlocal,zhou2020flat,zhou2020metasurface,huo2020photonic} 
Traditionally, polarization control has relied on contact-based devices such as waveplates, polarizers, and retarders, which apply uniform Jones matrix operations at a single physical interface. This direct-contact architecture inherently limits spatial programmability, design flexibility, and applicability in environments incompatible with physical structures, such as high-power, vacuum, or bio-sensitive systems. Recent advances in metasurfaces—planar arrays of subwavelength meta-atoms—have enabled fine control over amplitude, phase, and polarization of light at subwavelength resolution. By tailoring the geometry and orientation of individual meta-units, metasurfaces can implement spatially varying Jones matrix operations across the device plane. However, the underlying Jones matrix operations remain confined to the physical interface of the metasurface itself.\cite{li2024optical,chen2016review}

To overcome this limitation, recent approaches have sought to extend Jones matrix operations into free space. For example, propagation-dependent metasurfaces have been developed to modulate polarization longitudinally along the optical axis, enabling tailored Jones matrix transformations at different propagation distances.\cite{dorrah2021metasurface} Separately, Jones-matrix holography has been explored, where the far-field diffraction pattern of a metasurface hologram is engineered to encode a spatially varying polarization response.\cite{rubin2021jones} Additional innovations, such as compact full-Stokes polarization cameras implemented by metasurface gratings,\cite{rubin2019matrix} further demonstrate the promise of non-contact, propagation-based polarization engineering.

In this work, we introduce a fundamentally new concept in polarization optics: action-at-a-distance polarization control using virtual polarization elements (VPEs). Our framework builds on a high-dimensional plane wave superposition approach, where a metasurface at the modulation plane physically implements the designed polarization modulation enabling deterministic realization of prescribed Jones matrix transformations 
\(\tilde{J}(x,y)\) across a target plane in free space. Crucially, this target plane is physically contactless and material-free, decoupled from the metasurface itself, allowing collective synthesis of the Jones matrix across multiple polarization elements instead of relying on one-to-one local mappings.

We demonstrate this capability through three representative implementations: single-function VPEs that uniformly realize circular polarizer (CP), \( 45^\circ \) linear polarizer (LP), half-wave plate (HWP), and quarter-wave plate (QWP) operations; a multifunction VPE that simultaneously encodes spatially distinct Jones matrix functions with arbitrary relative phase offsets; and VPE-based vortex waveplates that generate structured vector vortex beams via spatially varying HWP operations. We envision this work as a foundation for emerging non-contact and nonlocal polarization architectures.

\section{Design concept}
In this section, we present the design concept of VPEs, which enable the realization of arbitrary polarization operations defined by a spatially varying Jones matrix distribution \( \tilde{F}(x,y) = e^{i\varphi(x,y)} \cdot \tilde{J}(x,y) \) on a designated free-space plane at \( z = z_0 \) (Figure~\ref{fig:1}a). Here, \( \varphi(x,y) \) and \( \tilde{J}(x,y) = \left[\renewcommand{\arraystretch}{0.6} \begin{array}{cc}j_1 & j_2 \\j_3 & j_4\end{array} \right] \) represent the preset additional phase and Jones matrix, respectively. Throughout this paper, we use " \( \sim \) ", "\( |\cdot\rangle \)", and bold symbols to denote 2$\times$2 Jones matrices, 2$\times$1 polarization state vectors, and complex field amplitude, respectively. Physically, the polarization function \( \tilde{F}(x,y) \) can be convinced as superpositions of plane waves weighted by 2$\times$2 Jones matrices rather than scalar complex amplitudes .\cite{dorrah2021metasurface} The matrix-valued weighting factor for each plane wave can be determined via a high-dimensional Fourier transform:

\begin{equation}
\tilde{A}(k_x^u, k_y^v) = \frac{1}{L_x L_y} \int_0^{L_x} \int_0^{L_y} \tilde{F}(x,y) e^{-i(k_x^u x +  k_y^v y)} \, dx \, dy
\label{eqn1}
\end{equation}
\noindent 
where \( L_x \) and \( L_y \) define the rectangular region implementing \( \tilde{F}(x,y) \), \( k_x^u =2\pi u/{L_x} \) and \( k_y^v = 2\pi v/{L_y} \) are \( \hat{x} \)- and \( \hat{y} \)-direction  wave numbers, respectively, with integers \( u = -N_1 \dots 0 \dots N_1 \) and \( v = -N_2 \dots 0 \dots N_2 \). We exclude evanescent waves by imposing \( (k_x^u)^2 + (k_y^v)^2 \leq (2\pi/{\lambda})^2\) and choose positive \( \hat{z} \)-direction wavenumber \( k_z^{(u,v)}\) to ensure all waves propagate in positive \(\hat{z}\) direction. Consequently, the superposition of these \( (2N_1+1) \times (2N_2+1) \) high-dimensional plane waves, represented as \(\sum_{u=-N_1}^{N_1} \sum_{v=-N_2}^{N_2} \tilde{A}(k_x^u, k_y^v) e^{i(k_x^u x + k_y^v y)} e^{i k_z^{(u,v)} (z -z_0)}\), on the \( z = 0 \) plane yields:
\begin{equation}
 \tilde{M}(x,y) = \sum_{u=-N_1}^{N_1} \sum_{v=-N_2}^{N_2} \tilde{A}(k_x^u, k_y^v) e^{i(k_x^u x + k_y^v y)} e^{i k_z^{(u,v)} (-z_0)}
\label{eqn2}
\end{equation}
\noindent where \(\tilde{M}(x,y) \) is the required 2$\times$2 Jones matrix for the target function \(\tilde{F}(x,y)\). Once an optical device that can, point by point, implement \(\tilde{M}(x,y) \) on the \( z = 0 \) plane, the output field on the \( z = z_0 \) plane (i.e., \( |\mathbf{E_{\text{out}}}\rangle \)) and the incident field on the \( z = 0 \) plane (i.e., \(|\mathbf{E_{\text{in}} }\rangle\)) satisfy the relationship \( |\mathbf{E_{\text{out}}}\rangle \big|_{z=z_0} \approx \tilde{F}(x,y) \cdot  |\mathbf{E_{\text{in}} }\rangle\), thereby achieving the preset Jones matrix operation (Supporting Information, S1). Notably, the plane at which polarization conversion takes place is suspended in free space, free from any polarization devices or physical contact. This feature is especially beneficial for applications demanding non-contact polarization modulation.

\clearpage
\section{Metasurface implementation}
\begin{figure}[H]
  \centering
   \includegraphics[width=1\linewidth]{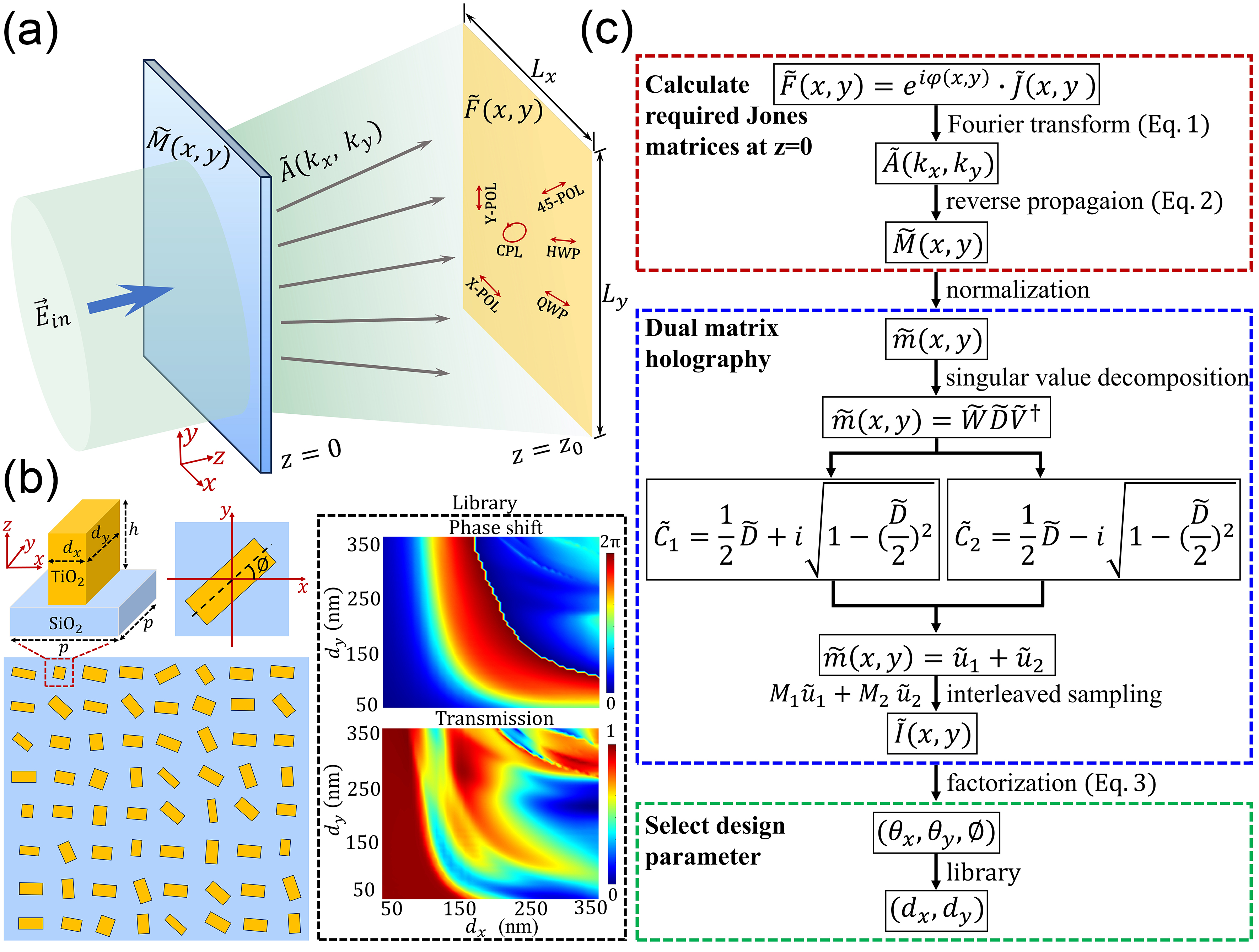} 
 \caption{The concept, design method, and implementation of VPE. (a) Schematic of the VPE, represented by \(\tilde{M}(x, y) \), a transverse distribution of 2$\times$2 Jones matrices on the plane \( z = 0 \), enabling diverse polarization operations \( \tilde{F}(x, y) \) on the free-space plane \( z = z_0 \). (b) Schematic of a TiO\(_2\) nanopillar metasurface. Phase and transmission responses as functions of (\(d_x,d_y\)) for \(x\)-polarization (\(h = 600 \: \)nm, \(p = 0.5 \:\mu\)m, \(\lambda=532 \: \)nm), serving as a library for metasurface implementation. (c) Design flowchart of the VPE: First, calculate \(\tilde{M}(x, y) \) by superimposing high-dimensional plane waves on the plane \( z = z_0 \). The weighting factors for these plane waves, \( \tilde{A}(k_x, k_y) \), are determined by \( \tilde{F}(x, y) \) through Fourier transforms. Next, obtain \( \tilde{I}(x, y) \) using double-matrix holography for the VPE implementation with a pure-phase metasurface. Finally, retrieve the design parameters (\(d_x,d_y,\phi\)).}
  \label{fig:1}
\end{figure}

In principle, any birefringent optical elements, such as the pixels of a spacial light modulator or  unit cells of a nanopillar metasurface,\cite{balthasar2017metasurface,arbabi2015dielectric,kruk2016invited,shi2020continuous} that can realize \(\tilde{M}(x, y)\) at each \((x, y)\) positions are applicable in implementation the proposed VPEs. This work adpots the metasurface scheme by using fixed-height rectangular Titanium Dioxide (TiO\(_2\)) nanopillars atop a glass substrate (Figure~\ref{fig:1}(b)).\cite{devlin2016broadband} Due to anisotropy, these nanopillars support two propagating modes that experiencing different phase delays, realizing a local Jones matrix of the form:\cite{saleh2019fundamentals,escuti2016controlling}
\begin{equation}
\tilde{I}(x, y) = R\left(-\phi(x, y)\right)
\left[\renewcommand{\arraystretch}{0.6}
\begin{matrix}
e^{i\theta_x(x, y)} & 0 \\
0 & e^{i\theta_y(x, y)}
\end{matrix}
\right]
R\left(\phi(x, y)\right)
\label{eqn3}
\end{equation}
\noindent where \((\theta_x,\theta_y)\) are phase retardances tuned by the nanopillar's dimensions (\(d_x,d_y\)), \(R\left(\phi(x,y)\right) = \left[\renewcommand{\arraystretch}{0.6}\begin{matrix}\cos\phi & \sin\phi \\\sin\phi & \cos\phi\end{matrix}\right]\) is the rotation matrix with \( \phi \) being the nanopillar's in-plane rotational angle.\cite{arbabi2015dielectric,balthasar2017metasurface} Since \(\tilde{I}(x, y)\) is symmetric and unitary, it may not be able to implement \( \tilde{M}(x, y)\) derived from \ Eq.~\ref{eqn2}. The symmetry condition can be satisfied by setting a symmetric matrix \(\tilde{F}(x, y) \). The unitary issue can be addressed by employing the dual-matrix holography technique,\cite{mendoza2014encoding,hsueh1978computer} which transforms \( \tilde{M}(x, y) \) into the form of Eq.~\ref{eqn3} in following steps: (i)\:The matrix \(\tilde{M}(x,y)\) is normalized to \(\tilde{m}(x,y)\); (ii)\:Singular value decomposition of \(\tilde{m}(x,y)=\tilde{W}\tilde{D}\tilde{V}^\dagger\), where \( \tilde{D} \) is a diagonal matrix with non-negative real singular values, \( \tilde{W} \) and \(\tilde{V} \) are unitary matrices, and "\(\dagger\)" denotes the conjugate transpose; (iii)\:Decompose \( \tilde{D} \) into the sum of two matrices with unit determinant, thus, \( \tilde{m} \) is decomposed into the sum of the two unitary matrices; (iv)\:Using checkerboard masks \( M_1 \) and \( M_2 \) (\( M_1 + M_2 = 1 \)) to perform complementary sampling of \( \tilde{u}_1 \) and \( \tilde{u}_2 \), yielding \( \tilde{I}(x,y) \). Then, the metasurface can be implemented by determining the optimal nanopillar geometry (\(d_x,d_y,\phi\)) that defines the
transformation matrix with the smallest deviation from the obtained \(\tilde{I}(x, y)\).\cite{dorrah2021metasurface} The design flowchart of the VPE is summarized in Figure \ref{fig:1}(c). 

\clearpage
\section{Results and discussion}

\subsection{Single-function VPEs}
\begin{figure}[H]
  \centering
 \includegraphics[width=0.88\linewidth]{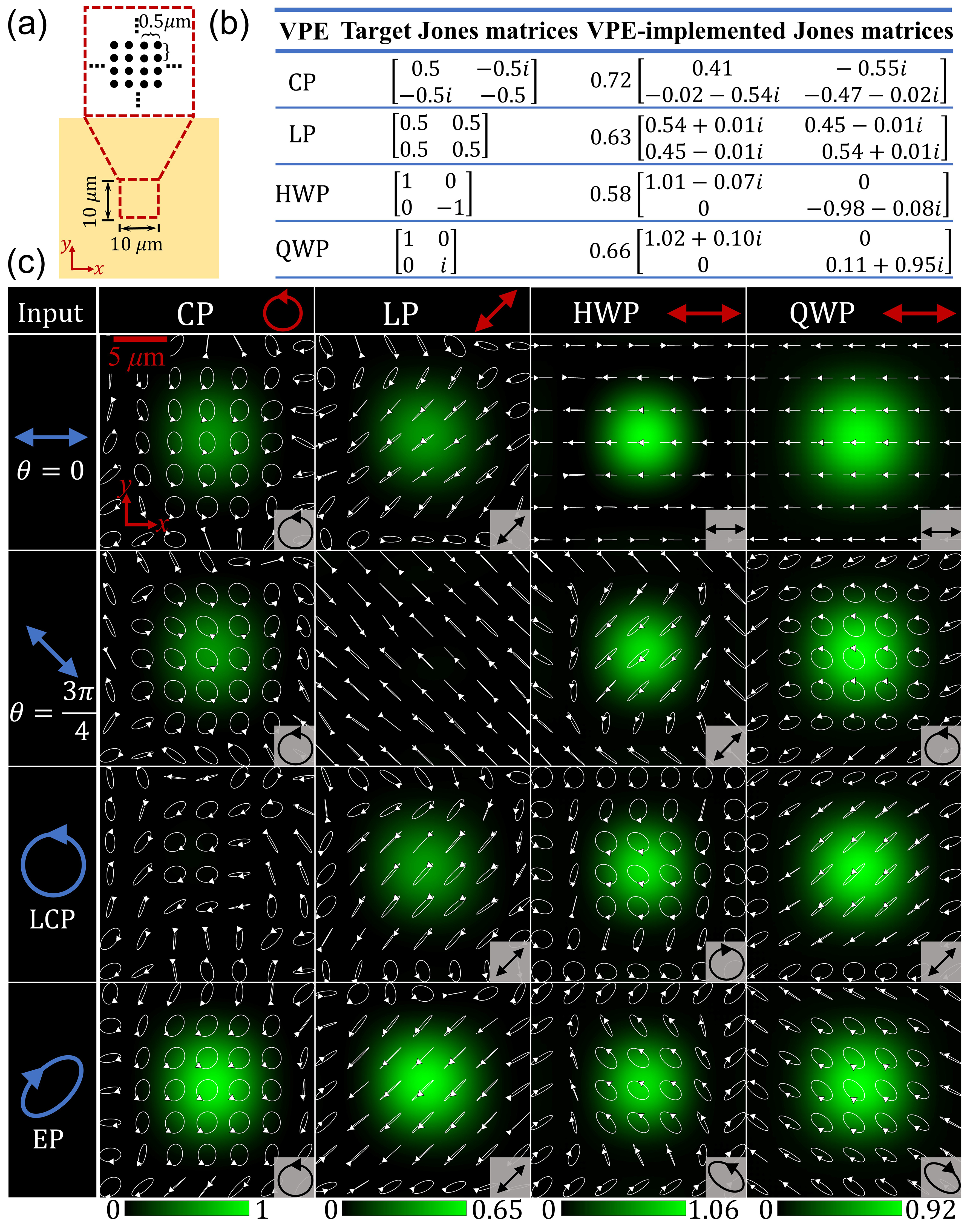} 
\caption{Single-function VPEs performing a uniform polarization operation. (a) Assigned Jones matrix over a \( 10 \,\mu\text{m} \times 10 \,\mu\text{m} \) central region on the target plane, discretized with a \( 0.5 \,\mu\text{m}\) grid. (b) Comparison between target and VPE-realized Jones matrices. (c) Normalized intensity (relative to incident light) and polarization distribution under various incident polarizations: linear at \( \theta = 0 \) (first row) and \( \theta = 3\pi/4 \) (second row), left-handed circular (third row), and elliptical (four row). White ellipses with directional arrows indicate local polarization and instantaneous phase. Insets: polarization predicted by target Jones matrix.}
  \label{fig:2}
\end{figure}

We begin with four single-function VPEs, each implementing a distinct Jones matrix \(\tilde{J}\):
\(\renewcommand{\arraystretch}{0.6}\begin{bmatrix} 0.5 & -0.5i \\ -0.5i & -0.5 \end{bmatrix},
\renewcommand{\arraystretch}{0.6}\begin{bmatrix} 0.5 & 0.5 \\ 0.5 & 0.5 \end{bmatrix}, \renewcommand{\arraystretch}{0.6}\begin{bmatrix} 1 & 0 \\ 0 & -1 \end{bmatrix}, \renewcommand{\arraystretch}{0.6}\begin{bmatrix} 1 & 0 \\ 0 & i \end{bmatrix}\), corresponding to a left-handed CP, a LP at \( 45^\circ \) to the horizontal, a HWP with fast axis horizontal, and a QWP with fast axis horizontal, respectively. To comply with the symmetric matrix requirement for \(\tilde{I}(x,y)\), a modified CP matrix is adopted to convert linear input  into left-handed circular polarization while suppressing left-handed circular input.\cite{rubin2021jones} The operation function \(\tilde{F}(x, y) \) is defined over a \( L_x \times L_y = 60\,\mu\text{m} \times 60\,\mu\text{m} \) domain on the plane \( z = z_0 = 10\,\mu\text{m} \), taking nonzero values only within a central \( 10\,\mu\text{m} \times 10\,\mu\text{m} \) region. There, \(\tilde{F}(x, y) \) is assigned \(\tilde{J} \) on a uniform \( 0.5\,\mu\text{m} \) grid, formally written as \(\tilde{F}(x, y) = \sum\limits_{m,n=-10}^{10} \tilde{J} \, \delta(x - 0.5m, y - 0.5n) \) (Figure~\ref{fig:2}a). The metasurface is fixed at \(60\,\mu\text{m} \times 60\,\mu\text{m}\), with \(N_1=N_2=4\) in Eq. 1. 

The optical response of the VPEs is evaluated by performing three-dimensional finite-difference time-domain (FDTD) simulations to compute the transmitted field on the metasurface exit plane, followed by free-space propagation to the target plane using the vectorial angular-spectrum representation.\cite{carter1972electromagnetic,deng2007analytical} The implemented Jones matrix, denoted as \(\tilde{J'}\), was retrieved by: (i) calculating the output fields \(\mathbf{E}_{\text{out}}|_{x}\) and \(\mathbf{E}_{\text{out}}|_{y}\), under unit-amplitude \(x\)- and \(y\)-polarized plane-wave incidence; (ii) spatially averaging these fields over the designated \(10 \ \mu \)m \(\times\) \(10 \ \mu \)m region to obtain \(\mathbf{\overline{E}}_{\text{out}}|_{x} =\bar{\mathbf{E}}_x|_x \hat{x} + \bar{\mathbf{E}}_y|_x\hat{y}\) and \(\mathbf{\overline{E}}_{\text{out}}|_{y} =\bar{\mathbf{E}}_x|_y \hat{x} + \bar{\mathbf{E}}_y|_y\hat{y}\); (iii) reconstructing the Jones matrix as \(\left[\renewcommand{\arraystretch}{0.6} \begin{array}{cc}\bar{\mathbf{E}}_x|_x  & \bar{\mathbf{E}}_x|_y  \\\bar{\mathbf{E}}_y|_x  & \bar{\mathbf{E}}_y|_y \end{array} \right]\). To facilitate comparison with the target Jones matrix \(\tilde{J} \), \(\tilde{J'}\) is written in the form \( Ae^{i\varphi}
\renewcommand{\arraystretch}{0.6}\begin{bmatrix} a & b \\ c & d \end{bmatrix}\), where \(A\), \(\varphi\), and \((a, b, c, d)\) are determined by minimizing its Frobenius distance to \(\tilde{J} \).\cite{horn2012matrix,golub2013matrix} 

Figure~\ref{fig:2}b compares the implemented Jones matrices \(\tilde{J'} \) of the four VPEs with their target matrices 
\(\tilde{J} \), showing close agreement apart from a global amplitude factor that does not affect the polarization transformation. The closeness between the realized and target Jones matrices was evaluated using matrix  fidelity, defined as  \(|\mathrm{Tr}(\tilde{J}^\dagger \tilde{J}')| / (\mathrm{Tr}(\tilde{J}^\dagger \tilde{J}')\mathrm{Tr}(\tilde{J'}^\dagger \tilde{J}))^{1/2} \) denotes the matrix trace. The computed fidelities of 0.993, 0.995, 0.997, and 0.994 for CP, LP, HWP, and QWP, respectively, confirm the high accuracy of the implemented polarization operations. Minor deviations arise from limitations of the nanopillar library, where geometry (\(d_x,d_y)\) cannot exactly match the local \(\tilde{I}(x, y)\) (Supporting Information, S3), which could be mitigated by improving nanopillar transmittance via antireflection coatings or impedance-matching design.\cite{devlin2016broadband} To further evaluate performance, we analyze the output intensity and polarization distributions under various input polarization states (Figure~\ref{fig:2}c). Two quantitative metrics are used: (i) Polarization fidelity, \(
\eta = \left. \iint |\mathbf{E}_{\text{desired}}|^2 \,dx\,dy \middle/ \iint |\mathbf{E}_{\text{out}}|^2 \,dx\,dy \right.
\), representing the energy ratio of the desired polarization state \( \mathbf{E}_{\text{desired}} \) and the output field \( \mathbf{E}_{\text{out}} \); (ii) Conversion efficiency, \(
C = \left. \iint |\mathbf{E}_{\text{desired}}|^2 \,dx\,dy \middle/ \text{total incident energy} \right.
\), quantifying the fraction of converted energy. All integrals are evaluated over the designated \(10\,\mu\text{m} \times 10\,\mu\text{m}\) region. All single-function VPEs exhibit high-fidelity performance (\(\eta\) \(>\) 0.975) with \(C\) reaching approximately 55\% (CP), 38\% (LP), 39\% (HWP), and 46\% (QWP) of the theoretical maximum (Supporting Information, S4). The relatively low conversion efficiencies stem from the limited design area allocated to each function. Expanding this designated region markedly enhances performance; for example, simulations indicate that increasing the area to \(60\,\mu\text{m} \times 60\,\mu\text{m}\) boosts \(C\) to 0.75, 0.67, 0.71, and 0.68 for CP, LP, HWP, and QWP, respectively.

\subsection{Multifunction VPE}
\begin{figure}[H]
  \centering
 \includegraphics[width=0.88\linewidth]{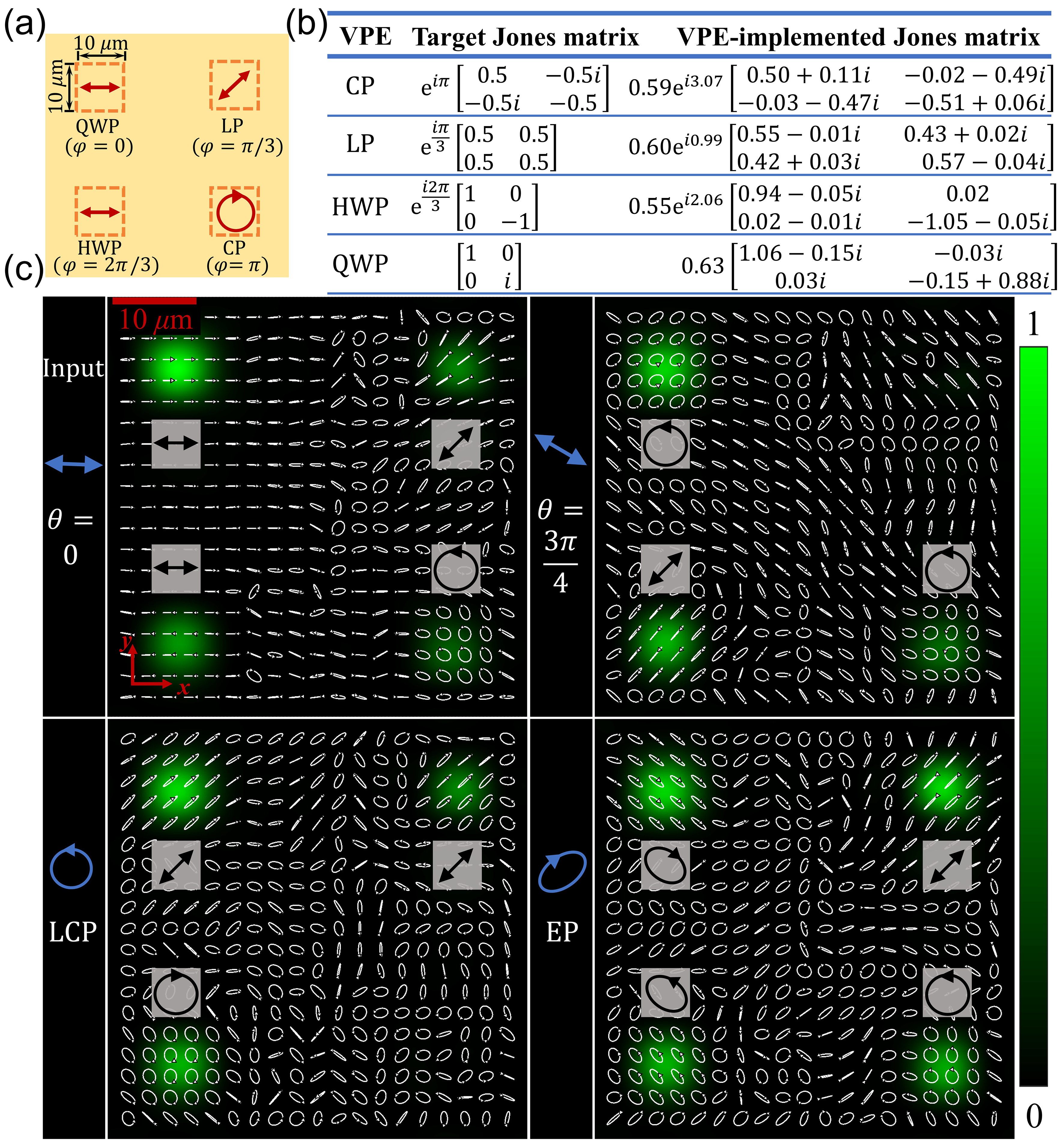} 
\caption{Multi-function VPE enabling region-specific polarization operations with prescribed relative phase. (a) Assigned Jones matrices over four \( 10 \,\mu\text{m} \times 10 \,\mu\text{m} \) regions on the target plane. (b) Comparison between target and VPE-realized Jones matrices. (c) Intensity (normalized to incident light) and polarization distribution under various incident polarizations: linear at \(\theta = 0\) (top-left) and \(\theta = 3\pi/4\) (top-right), left-handed circular (bottom-left), and elliptical (bottom-right). White ellipses with directional arrows indicate local polarization and instantaneous phase. Insets: polarization predicted by target Jones matrix.}
  \label{fig:3}
\end{figure}

\justifying
Owing to its design flexibility, we further showcase a multifunction VPE that implements distinct Jones matrices with independently assigned additional phases \( \varphi \) across four  \(10\,\mu\text{m} \times 10\,\mu\text{m}\) regions centered at \((\pm20\ \mu\text{m},\pm20 \ \mu\text{m})\) (Figure~\ref{fig:3}a). The target matrices are: \\ \(e^{i\pi} \left[\renewcommand{\arraystretch}{0.6} \begin{matrix} 0.5 & -0.5i \\ -0.5i & -0.5 \end{matrix} \right]\), \(e^{i\pi/3} \left[\renewcommand{\arraystretch}{0.6} \begin{matrix} 0.5 & 0.5 \\ 0.5 & 0.5 \end{matrix} \right] \), \(e^{i2\pi/3} \left[\renewcommand{\arraystretch}{0.6} \begin{matrix} 1 & 0 \\ 0 & -1 \end{matrix} \right]\),  \(\left[\renewcommand{\arraystretch}{0.6} \begin{matrix} 1 & 0 \\ 0 & i \end{matrix} \right]\), corresponding to a left-handed CP \( \varphi = \pi \), a \( 45^\circ \) LP \( \varphi = \pi/3 \), a HWP with horizontal fast axis \( \varphi = 2\pi/3 \), and a QWP with horizontal fast axis \( \varphi = 0 \), respectively. All other parameters are consistent with those in the single-function case.

Similar to their single-function counterparts, the multifunction VPE achieves intended polarization operations with matrix fidelities of 0.992, 0.990, 0.997, and 0.983 for CP, LP, HWP, and QWP, respectively (Figure~\ref{fig:3}b). Regarding the additional phases, after referencing to the QWP region, the realized relative phases are 0, 0.99, 2.06, and 3.07 radians, closely matching the target values of 0, \(2\pi/3\), \(\pi/3\), and \(\pi\), with relative errors below 6\%. Figure~\ref{fig:3}c displays the spatial intensity and polarization distributions under various input polarizations, demonstrating high-fidelity performance (\(\eta\) \(>\) 0.937) for all sub-elements, with conversion efficiencies \(C\) reaching 36\% (CP), 39\% (LP), 33\% (HWP), and 41\% (QWP) of the theoretical limit (Supporting Information, S4), thereby confirming the ability to simultaneously implement distinct Jones matrices with precise phase control on a free-space plane.  

Notably, the ability to impose distinct Jones matrices across neighboring regions is fundamentally limited by the system’s spatial frequency bandwidth and material constraints. Specifically, the maximum transverse wavevector \(k_r^{\max} = \sqrt{k_x^2 + k_y^2}\) sets the minimum resolvable feature size \(\Delta x_{\min} \approx \pi/{k_r^{\max}}\), beyond which sharp Jones matrix variations cannot be faithfully reproduced. Moreover, the achievable modulation is bounded by the material response—namely, the tunable range, fabrication precision, and optical losses of the metasurface elements—placing practical limits on the spatial resolution and fidelity of multifunction VPEs.

\subsection{VPE-based vortex waveplate}
We further demonstrate the versatility of VPEs by implementing vortex waveplates,\cite{huang2019versatile, zhan2009cylindrical} enabling generation of vector vortex beams from a free-space plane through engineered distributions of HWP operations with spatially varying fast-axis orientations \((\theta)\) and additional phases \((\varphi)\). The spatial modulation follows \(
(\theta, \varphi) = \left( \frac{m}{2} \alpha + \theta_0, l \alpha + \varphi_0 \right)
\), where \(\alpha\) is the azimuthal angle; \(m\) and \(l\) are the polarization and phase topological charges; and \(\theta_0\), \(\varphi_0\) are initial offsets. We design four configurations on the \(z_0 = 10\,\mu\mathrm{m}\) plane: \((m, l) = (1, 0)\), \((4, 0)\), \((4, 1)\), and a dual-mode\cite{wang2007generation} combining \((1, 0)\) in the inner ring (\(r = 5\) to \(10\,\mu\mathrm{m}\)) and \((1, 0)\), and \((-2, 1)\) in the outer ring (\(r = 15 \: \text{to} \: 20\,\mu\mathrm{m}\)). Here, \(\theta_0 = \varphi_0 =  0 \), with other parameters kept the same as in the single-function design.

\begin{figure}[H]
  \centering
   \includegraphics[width=0.88\linewidth]{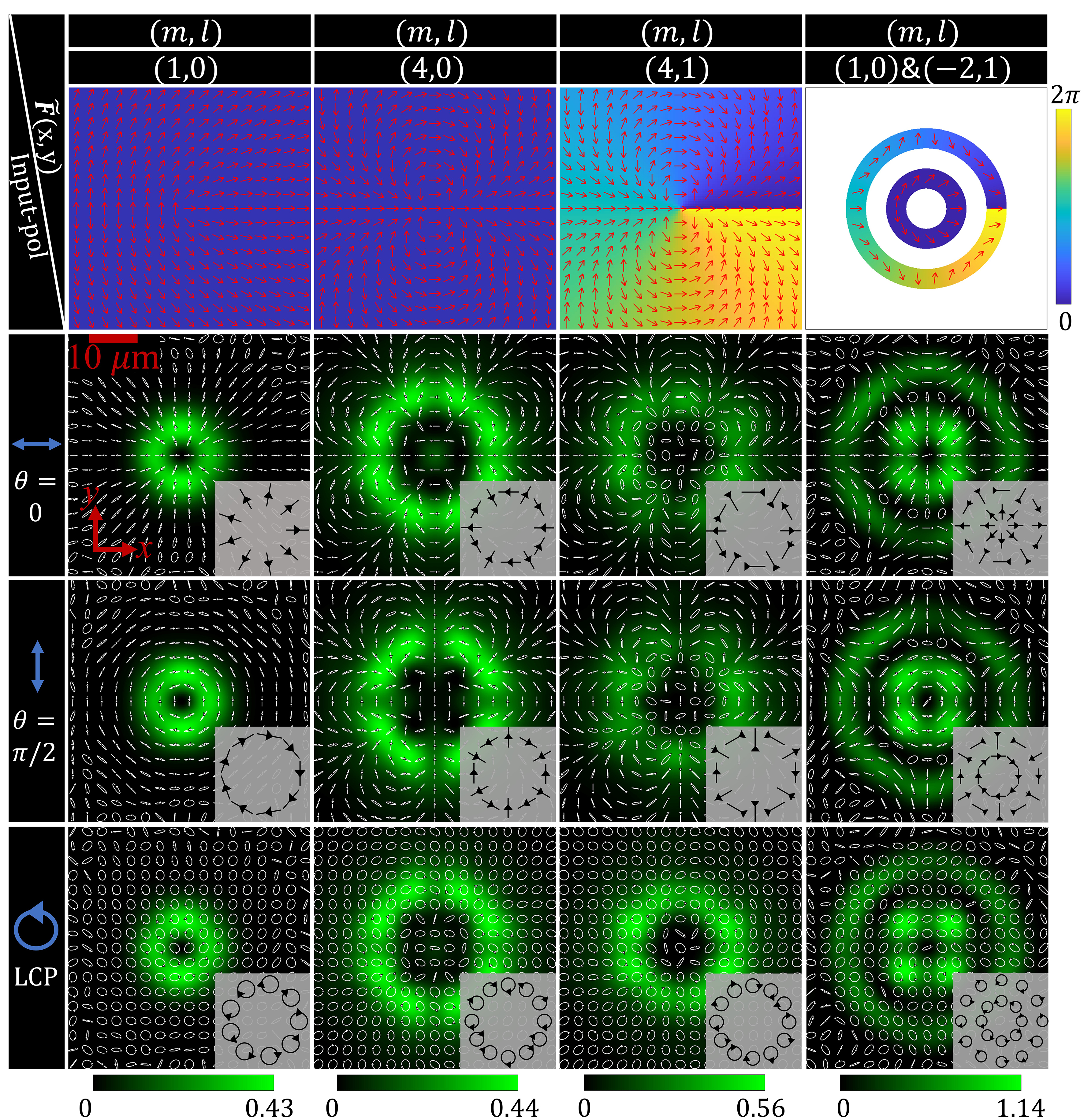} 
  \caption{VPE-based virtual vortex waveplates. Top row: Fast-axis orientations (red arrows) and additional phase shifts (colormap) of virtual HWPs arranged on a \( 0.5 \,\mu\text{m}\) grid over the target plane for topological charges \((m, l) = (1,0)\), \((4,0)\), \((4,1)\), and a dual-mode combination of \((1,0) \ \text{and} \ (-2,1)\). Row 2 to 4: Intensity (normalized to incident light) and polarization distribution under various incident polarizations, illuminated by Gaussian beams with waists of \(10 \ \mu\)m  (column 1), \(40 \ \mu\)m (columns 2\(\text{--}\)3), and \(50 \ \mu\)m (column 4). White ellipses with directional arrows indicate local polarization and instantaneous phase. Insets: polarization patterns predicted by target Jones matrix.}
  \label{fig:4}
\end{figure}

Figure~\ref{fig:4} highlights the capability of virtual vortex waveplates to generate distinct vector vortex beams under different incident polarizations. The $(m, l) = (1, 0)$ design yields radial or azimuthal polarization under \(x\)- and \(y\)-polarized illumination, respectively, and produces a vortex beam with opposite handedness and a spiral phase front  under left-handed circularly input, demonstrating spin-orbit interaction and central singularities in polarization or phase. \cite{bliokh2015spin} The $(4, 0)$ and $(4, 1)$ designs exhibit increasingly complex patterns, for both the fast-axis direction undergoes four rotations from 0 to \(\pi\) as the azimuthal angle \( \alpha \) spans from 0 to \( 2\pi \), with the latter imparts an additional azimuthal phase, resulting in spiral polarization structures. The dual-mode waveplate combines distinct topological orders in concentric regions, producing spatially multiplexed hybrid vector beams. These results confirm the versatility of VPEs in tailoring structured beams with region-specific polarization and phase control from a single metasurface.

\clearpage
\subsection{Conclusion}
In this work, we introduce a fundamentally new concept in polarization optics: action-at-a-distance polarization control enabled by VPEs. By exploiting a high-dimensional plane wave superposition framework, these devices impose a prescribed Jones matrix transformation between an input field at the modulation plane \((z=0\)), where a metasurface physically implements the required modulation, and an output field at the target plane (\(z=z_0\))—a contactless, material-free plane in free space where the desired polarization response occurs. We demonstrate four single-function VPEs that each implement the Jones matrices for CP, LP, HWP, and QWP operations; a multifunction VPE that simultaneously achieves these functions with arbitrary additional phase differences across distinct regions; and VPE-based vortex waveplates that generate structured vector vortex beams by spatially arranging HWP operations with continuously varying fast-axis orientations and phase profiles. While our implementation employs TiO\(_2\)
metasurfaces at wavelength of 532 nm, the concept applies broadly to other wavelength/platform capable of producing the required polarization modulation \(\tilde{I}(x,y)\) at \(z=0\). Although VPEs were designed at \(z_0=10 \ \mu\)m, the achievable distance is fundamentally limited only by the modulation plane's lateral dimension and spatial frequency content; for instance, we validated operation at \(z_0=300 \ \mu\)m by enlarging the modulation plane (Supporting Information, S5). Notably, placing the target Jones matrix on a contactless free-space plane—rather than directly at the exit surface of the polarization element—substantially enhances the flexibility and precision of polarization control by enabling collective synthesis of the Jones matrix across multiple polarization elements (Supporting Information, S6). The unique non-contact Jones matrix control offered by our VPE framework enables applications beyond the reach of conventional polarization devices, including remote polarization shaping in high-power laser systems, contamination-free biomedical operation, lossless control in quantum optics, and contactless calibration or encryption in free-space polarization systems.

\subsection*{\textbf{ASSOCIATED CONTENT}}
\noindent\textbf{Supporting Information}

\noindent The Supporting Information is available free of charge at https://.

\noindent\hspace{2em} Proof of the relation \( |\mathbf{E_{\text{out}}}\rangle \big|_{z=z_0} \approx \tilde{F}(x,y) \cdot |\mathbf{E_{\text{in}} }\rangle \); spectral filtering conditions; performance under an idealized nanopillar library; polarization fidelity \(\eta\) and conversion efficiency \(C\); extended-range multifunction VPE at \( z_0 = 300 \, \mu \text{m} \); non-contact vs contact designs: broader and more accurate Jones matrix realization (PDF)

\subsection*{\textbf{AUTHOR INFORMATION}}

\noindent\textbf{Corresponding Author}

\textbf{Jiayuan Wang} \hspace{0.05em} \rule[0.5ex]{0.3cm}{0.8pt} \hspace{0.05em} Department of Physics, Xiamen University, Xiamen 361005, China; 

\hspace{1em}Department of Shenzhen Research Institute of Xiamen University, Shenzhen 518057, 

\hspace{1em}China; orcid.org/0000-0003-0588-8689; Email: wangjiayuan@xmu.edu.cn

\noindent\textbf{Author}

\textbf{Mingyue Wang} \hspace{0.05em} \rule[0.5ex]{0.3cm}{0.8pt} \hspace{0.05em} Department of Physics, Xiamen University, Xiamen 361005, China

\subsection*{Author Contributions}
J. W. conceived the idea and supervised the project; M. W. implemented the simulations and performed the numerical analysis; M. W. and J. W. wrote the manuscript.

\noindent\textbf{Notes} 

\noindent The authors declare no competing financial interest.

\begin{acknowledgement}

This study was supported by the Guangdong Basic and Applied Basic Research Foundation (No.2021A1515011198).

\end{acknowledgement}
\bibliography{achemso-demo}

\clearpage

\begin{center}
\Huge\textbf{\underline{Supporting Information}}
\end{center}

\counterwithout{figure}{section}
\counterwithout{table}{section}
\counterwithout{equation}{section}

\setcounter{figure}{0}
\setcounter{table}{0}
\setcounter{equation}{0}

\renewcommand{\figurename}{Fig.}  
\renewcommand{\thefigure}{S\arabic{figure}}  

\renewcommand{\figurename}{Figure}  

\renewcommand{\thefigure}{S\arabic{figure}}

\captionsetup[figure]{labelfont=bf, labelsep=period, textfont=normalfont}  

\renewcommand{\thetable}{S\arabic{table}}  

\makeatletter
\def\tagform@#1{\maketag@@@{S(\ignorespaces#1\unskip\@@italiccorr)}}
\makeatother

\noindent\textbf{S1. Proof of the relation \( |\mathbf{E_{\text{out}}}\rangle  \big|_{z=z_0} \approx \bm{\tilde}{F}\bm{(x,y)}\cdot |\mathbf{E_{\text{in}}}\rangle \)}

\noindent When an incident plane wave with Jones vector \( |\mathbf{E_{\text{in}}}\rangle \) passes through an optical device described by \(\tilde{M}(x,y) \) on the plane \( z = 0 \), the transmitted field at this plane is given by: \(|\mathbf{E_{\text{out}} }\rangle \big|_{z=0} = \tilde{M}(x,y) \cdot |\mathbf{E_{\text{in}}}\rangle.\) Substituting the expression of \(\tilde{M}(x,y) \) from Eq.\:2 yields:
\begin{equation}
|\mathbf{E_{\text{out}} }\rangle \big|_{z=0} = \tilde{M}(x,y) \cdot |\mathbf{E_{\text{in}}}\rangle = \sum_{u=-N_1}^{N_1} \sum_{v=-N_2}^{N_2} \left[ \tilde{A}(k_x^u, k_y^v) e^{i(k_x^u x + k_y^v y)} e^{ik_z^{(u,v)} (-z_0)} \right] \cdot |\mathbf{E_{\text{in}}}\rangle.
\label{eqnS1}
\end{equation}
This represents a superposition of \( (2N_1+1) \times (2N_2+1)\) high-dimensional plane waves, each weighted a Jones matrix coefficients \(\tilde{A}(k_x^u, k_y^v)\). As these waves propagate to an arbitrary plane at \(z\), the field becomes:
\begin{equation}
|\mathbf{E_{\text{out}}}\rangle|_z = \sum_{u=-N_1}^{N_1} \sum_{v=-N_2}^{N_2} \left[ \tilde{A}(k_x^u, k_y^v) e^{i(k_x^u x + k_y^v y)} e^{ik_z^{(u,v)} (z-z_0)} \right] \cdot |\mathbf{E_{\text{in}}}\rangle.
\label{eqnS2}
\end{equation}
At the target plane \( z = z_0 \), this reduces to 
\begin{equation}
|\mathbf{E_{\text{out}}}\rangle|_{z=z_0} = \sum_{u=-N_1}^{N_1} \sum_{v=-N_2}^{N_2} \left[ \tilde{A}(k_x^u, k_y^v) e^{i(k_x^u x + k_y^v y)} \right] \cdot |\mathbf{E_{\text{in}}}\rangle.
\end{equation}
Using the Fourier relation for \( \tilde{A}(k_x^u, k_y^v) \) from Eq.\:1:
\begin{equation}
\sum_{u=-N_1}^{N_1} \sum_{v=-N_2}^{N_2} \left[ \left( \frac{1}{L_x L_y} \int_0^{L_x} \int_0^{L_y} \tilde{F}(x,y) e^{-i(k_x^u x + k_y^v y)} \, dx \, dy \right) e^{i(k_x^u x + k_y^v y)} \right] \approx \tilde{F}(x,y),
\end{equation}
Thus, the output field at the \( z = z_0 \) plane can be approximated as:
\begin{equation}
|\mathbf{E_{\text{out}}}\rangle|_{z=z_0} \approx \tilde{F}(x,y) \cdot |\mathbf{E_{\text{in}}}\rangle.
\end{equation}

\newpage\noindent\textbf{S2. Spectral filtering conditions}

\noindent  Upon applying dual-matrix holography, multiple copies of the Fourier spectra of the target operation \( \tilde{F}(x, y) \) generated at the Fourier plane. These higher diffraction orders, as the trade-off for the freedom afforded by this method, can be spatially filtered. To ensure the accurate preservation of the desired plane waves after filtering, the filter range must satisfy the following conditions:
\(
\Delta_x \geq 2 \cdot \max(k_x^u), \: \Delta_y \geq 2 \cdot \max(k_y^u)
\).
This condition ensures that the filter region encompasses twice the maximum wave vector component, thereby preventing the loss of critical spectral components. To avoid spectral aliasing, the sampling frequency is must satisfy the Nyquist sampling theorem requirement, i.e., being at least twice the highest signal frequency. In this study, the sampling frequency is set to 2 \(\mu\)m\(^{-1}\), corresponding to the metasurface period \(p = 0.5\: \mu\)m.

\begin{figure}[h]  
    \centering  
    \includegraphics[width=0.8\textwidth]{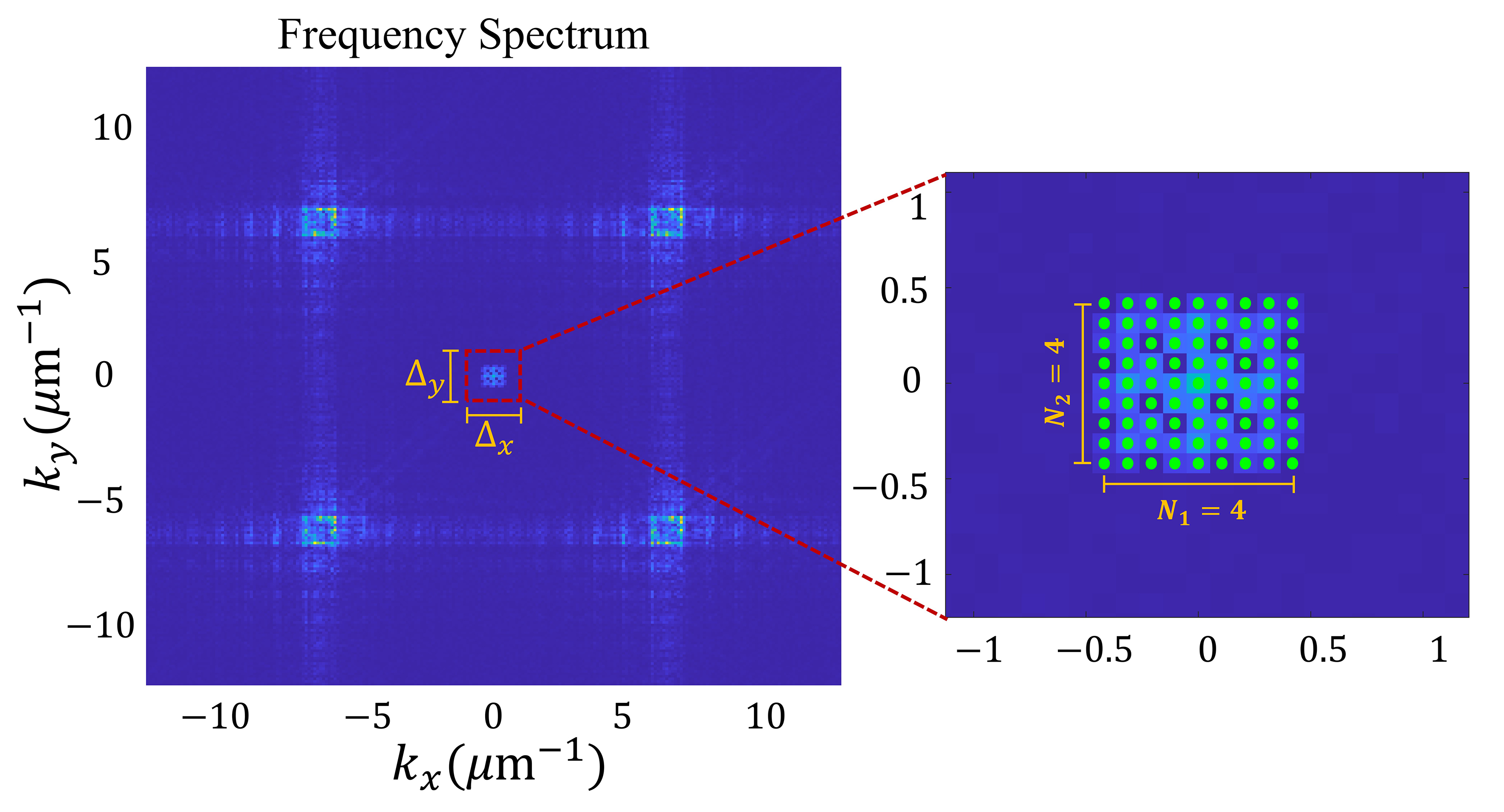}  
    \caption{Fourier space spectrum of the \( 45^\circ \) LP in Figure 2 of the main text. \( \Delta_x \) and \(\Delta_y \) indicate the filtering window (red box) at the spectrum center. Enlarged view of the filtering window reveals the chosen 81 plane waves (green dots).}  
    \label{fig:滤波图}  
\end{figure}

\newpage\noindent\textbf{S3. Performance under an idealized nanopillar library}

\begin{figure}[H]
  \centering
  \includegraphics[width=0.72\linewidth]{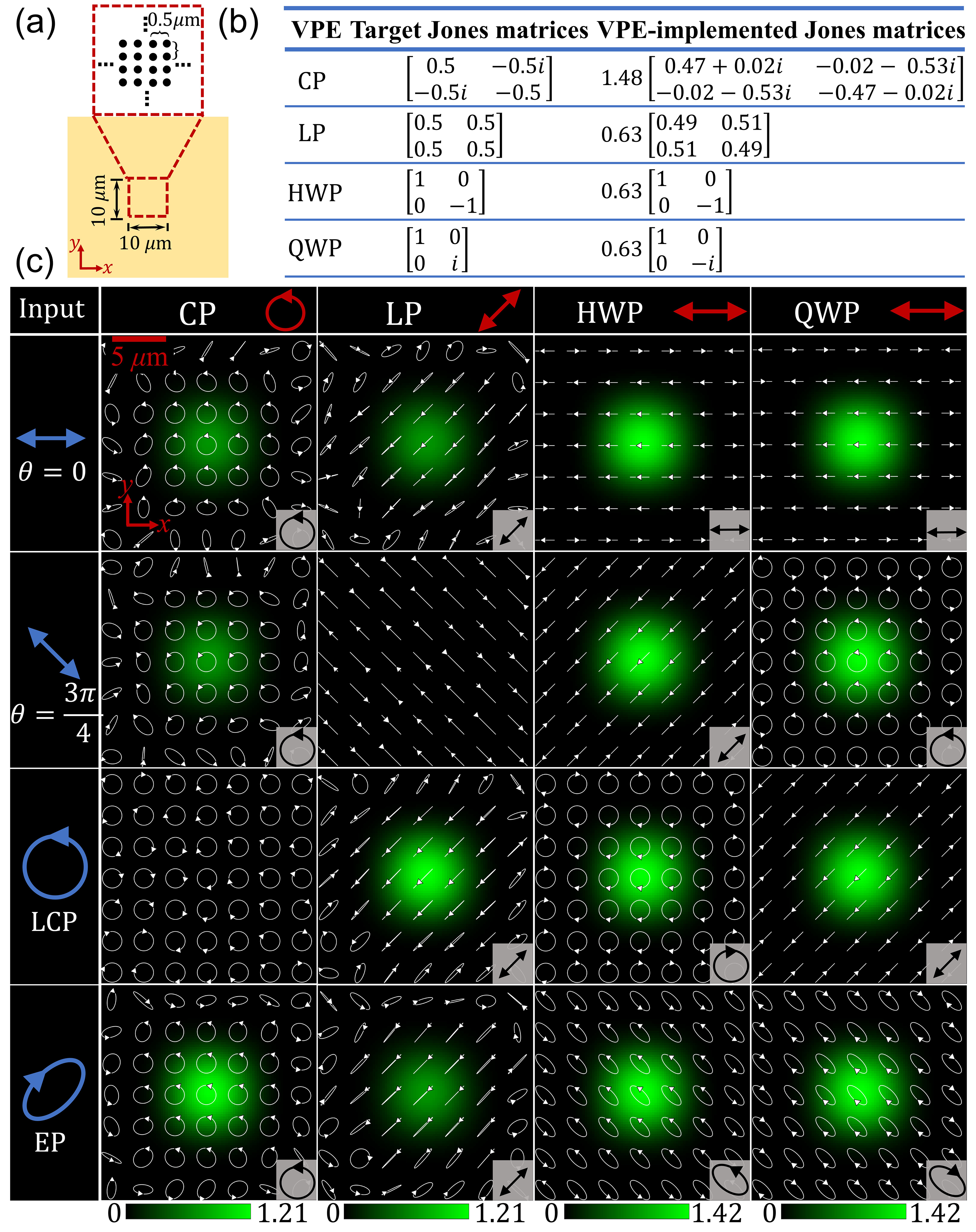} 
  \caption{Performance of single-function VPEs under an idealized nanopillar library, calculated using VAS method. (a) Assigned Jones matrix over a \( 10 \,\mu\text{m} \times 10 \,\mu\text{m} \) central region on the target plane. (b) Comparison between target and VPE-realized Jones matrices. Matrix fidelities are 0.997, 1, 1, and 1 for CP, LP, HWP, and QWP, respectively. (c) Normalized intensity (relative to incident light) and polarization distribution under various incident polarizations. White ellipses with directional arrows indicate local polarization and instantaneous phase. Insets: polarization predicted by target Jones matrix.}
  \label{fig:理想}
\end{figure}

\noindent  Using an idealized library, single-function VPEs achieve near-theoretical performance (Figure ~\ref{fig:理想}), confirming that deviations from target Jones matrices stem from practical nanopillar library limitations.

\clearpage

\clearpage
\noindent \textbf{S4. Polarization fidelity \(\eta\) and conversion efficiency \(C\)}

\noindent

\begin{figure}[H]
    \centering
    \includegraphics[width=0.68\textwidth]{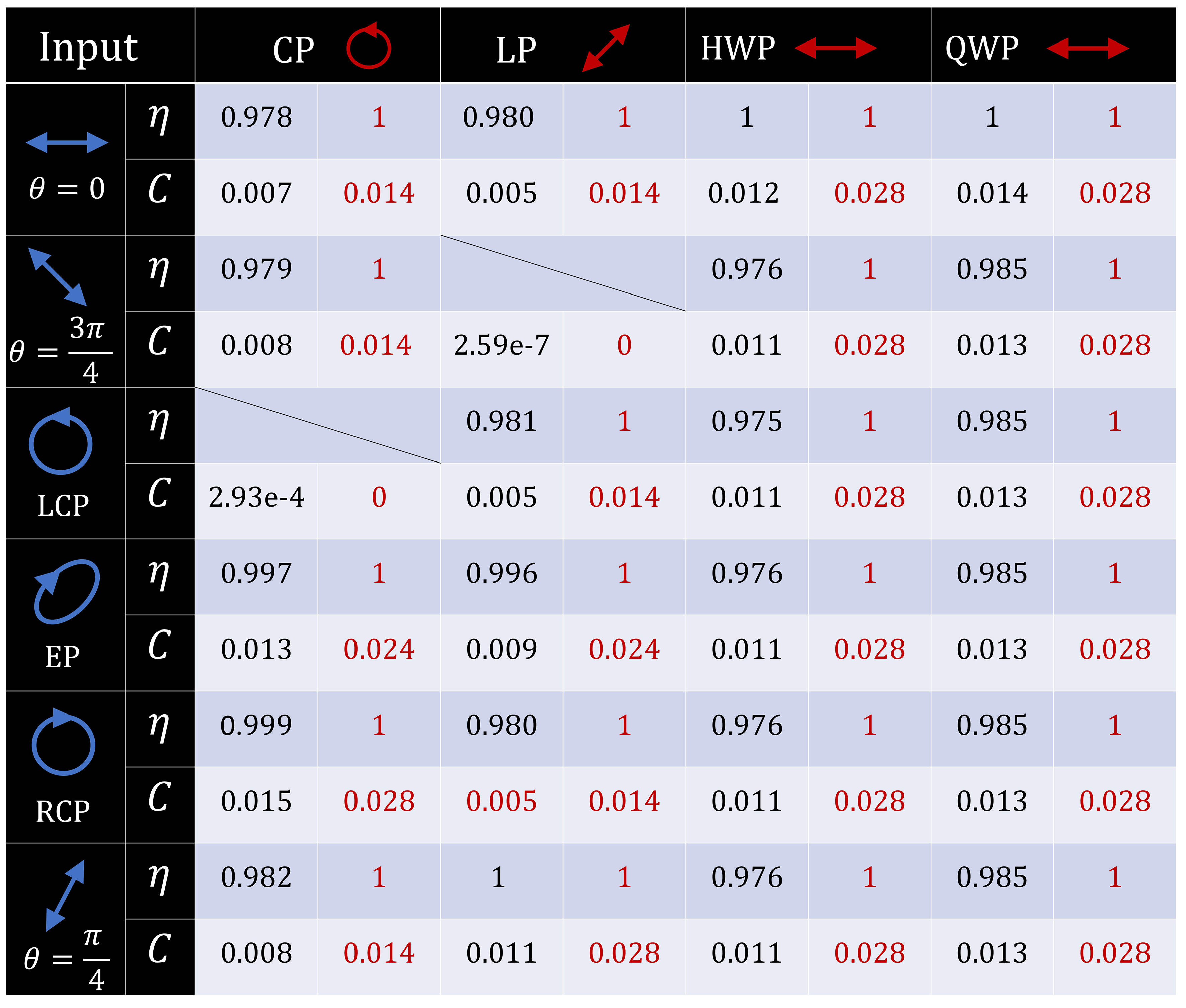}
     \captionsetup{type=table}  
    \caption{(\(\eta\), \(C\)) of single-function VPEs (black) vs ideal Jones matrix predictions (red).}
    \label{fig:图2(b)_转化效率}
\end{figure}

\begin{figure}[H]
    \centering
    \includegraphics[width=0.68\textwidth]{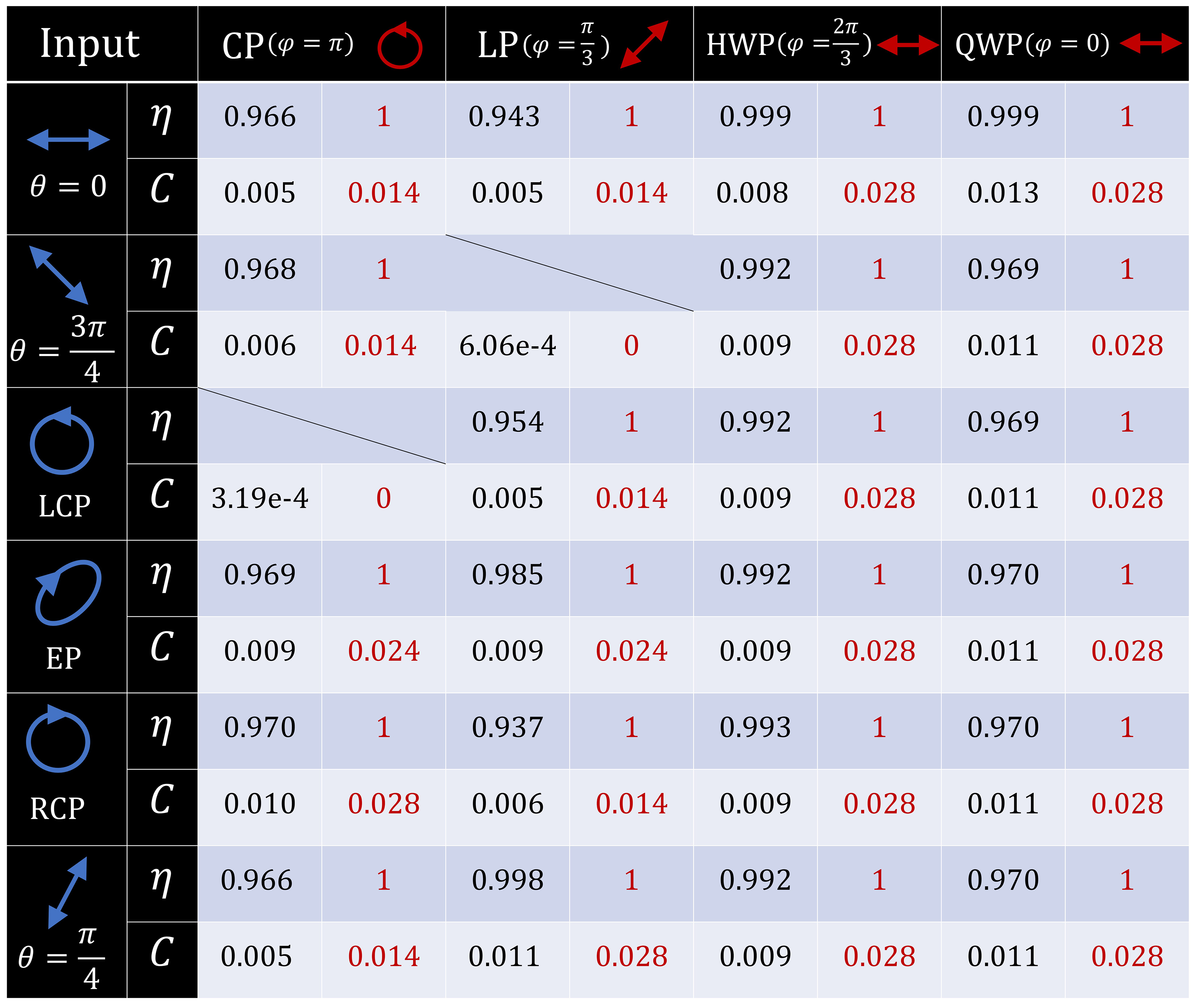} 
    \captionsetup{type=table}  
    \caption{(\(\eta\), \(C\)) of multifunction VPE (black) vs ideal Jones matrix predictions (red).}
    \label{fig:图3(b)_转化效率8}
\end{figure}

\newpage\noindent\textbf{S5. Extended-range multifunction VPE at \( z_0 = 300 \, \mu \text{m} \)}

\noindent To capture all spatial frequency components \(\tilde{A}(k_x,k_y)\), the metasurface must satisfy \(
D_x, D_y \geq L + \frac{\sqrt{2} z_0(k_r)_{max}}{\sqrt{((2\pi/\lambda)^2 - ((k_r)_{max})^2}}\), where \((k_r)_{\text{max}}\) is the maximum transverse wavevector \( k_r = \sqrt{k_x^2 + k_y^2} \).
For \( z_0 = 300\,\mu\text{m} \) and \( N_1 = N_2 = 5 \), this yields \(D_x=D_y \approx 200\,\mu\text{m}\).

\begin{figure}[H]
  \centering
\includegraphics[width=0.82\linewidth]{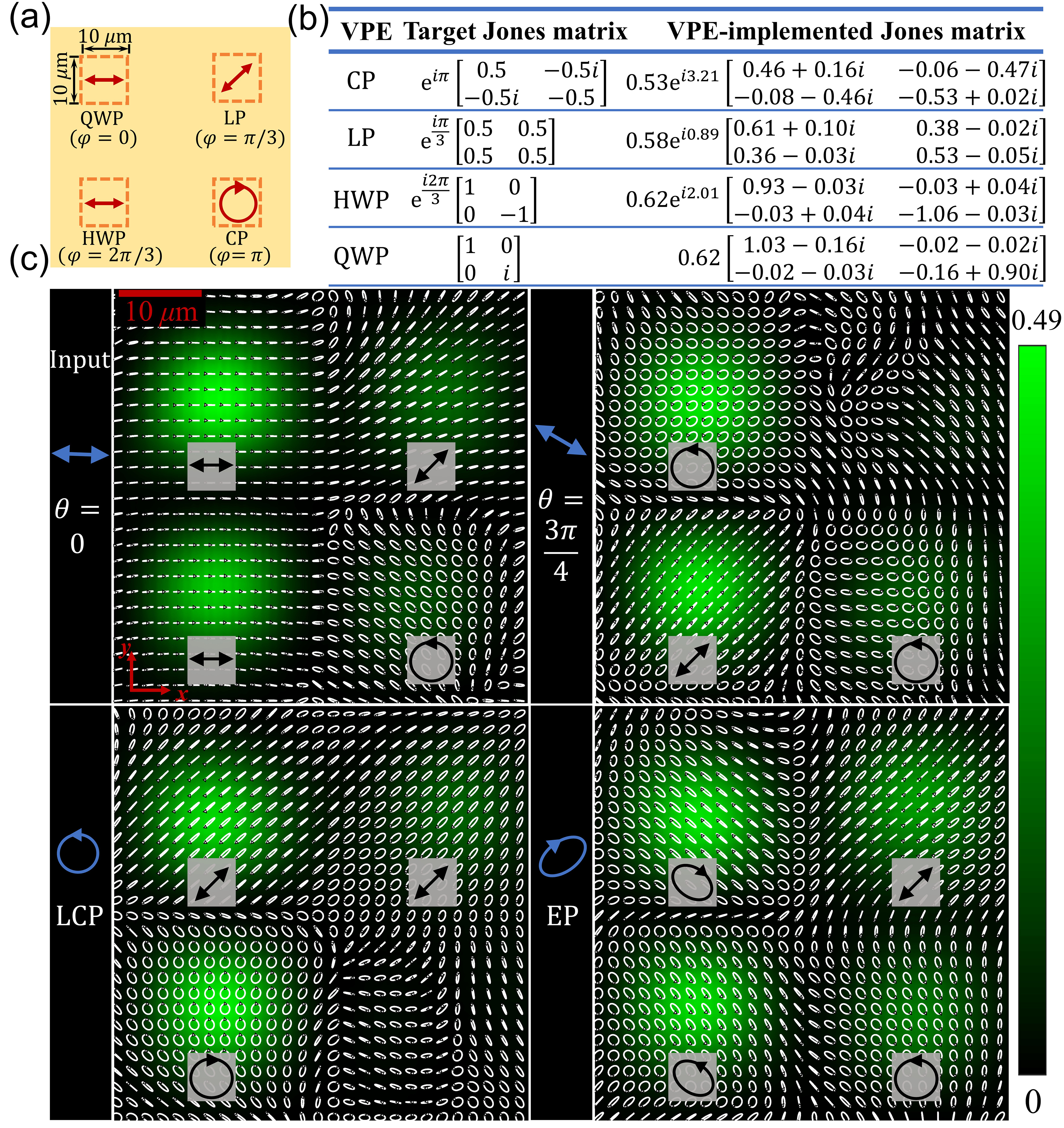} 
\caption{Multi-function VPE enabling region-specific polarization operations with prescribed relative phase on the plane \( z_0 = 300 \, \mu \text{m} \). The metasurface size is \( 200 \,\mu\text{m} \times 200 \,\mu\text{m} \), with \(N_1 = N_2 = 5\). (a) Assigned Jones matrices over four \( 10 \,\mu\text{m} \times 10 \,\mu\text{m} \) regions on the target plane. (b) Comparison between target and VPE-realized Jones matrices. Matrix fidelities are 0.979, 0.969, 0.996, and 0.985 for CP, LP, HWP, and QWP, respectively. (c) Intensity (normalized to incident light) and polarization distribution under various incident polarizations: linear at \(\theta = 0\) (top-left) and \(\theta = 3\pi/4\) (top-right), left-handed circular (bottom-left), and elliptical (bottom-right). White ellipses with directional arrows indicate local polarization and instantaneous phase. Insets: polarization predicted by target Jones matrix.}
  \label{fig:远距离衍射}
\end{figure}
\clearpage

\newpage\noindent\textbf{S6. Non-contact vs contact designs: broader and more accurate Jones matrix realization}

\noindent Placing the target Jones matrix on a free-space plane using a non-contact VPE, rather than directly on the exit plane of the element as in conventional contact designs, significantly improves the flexibility and accuracy of polarization transformations. This advantage arises from synthesizing the Jones matrix across multiple polarization elements instead of relying on a single local element. To illustrate this, we compare four polarization functions—CP, LP, HWP, and QWP—realized by the single-function VPE at \(z_0 = 10\,\mu\text{m}\), and a contact design, which corresponds to the limiting case where \(z_0 \rightarrow 0\), with the matrix governed by a single TiO\(_2\) nanopillar. Notably, no single-pillar solution exists for CP due to symmetry constraints, while for LP, HWP, and QWP, non-contact designs yield Jones matrices that closely match their target values.

\begin{figure}[H]
    \centering
    \includegraphics[width=1\textwidth]{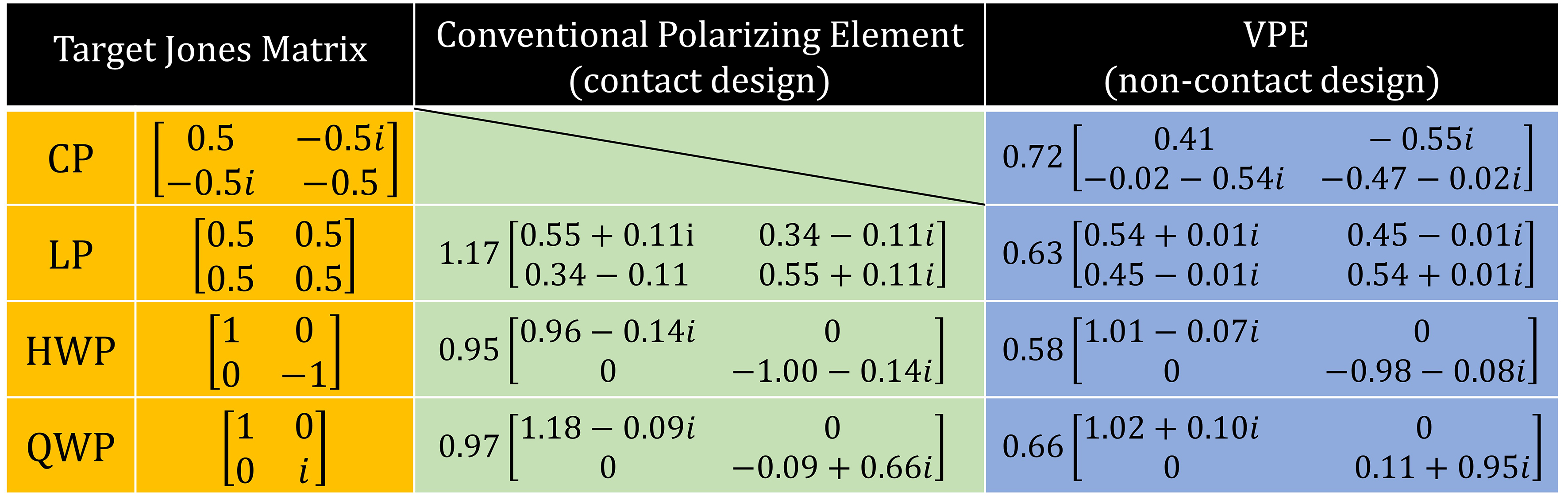} 
    \captionsetup{type=table}  
    \caption{Comparison of Jones matrix implementations using VPEs (non-contact design) and conventional polarization elements (contact design). Left column: target Jones matrices. Middle: matrices realized using single TiO\(_2\) nanopillars, with matrix fidelities of 0.944, 0.989, and 0.959 for the LP, HWP, and QWP, respectively. Right column: matrices realized by VPEs at \(z_0 = 10\,\mu\text{m}\), with matrix fidelities of 0.993, 0.995, 0.997 and 0.997 for the CP, LP, HWP, and QWP, respectively. Nanopillar library: \((d_x,d_y) \in [50\:\text{nm},350\:\text{nm}]\), \(h = 600 \: \)nm, \(p = 0.5 \:\mu\)m.}
    \label{fig:单米砖}
\end{figure}

\end{document}